\documentclass[aps,twocolumn,showpacs,preprintnumbers,amsmath,amssymb,nofootinbib,showkeys]{revtex4-1}
\usepackage{graphicx}
\newcommand{\amu}{a_{\mu}}

\usepackage{color}

\begin{document}

\title{Ballpark prediction for the hadronic light-by-light contribution to the muon $(g-2)_{\mu}$}

\author{Pere Masjuan} \email{Corresponding author: masjuan@kph.uni-mainz.de}
\author{Marc Vanderhaeghen}
\affiliation{Institut f\"ur Kernphysik, Johannes Gutenberg-Universit\"at, D-55099 Mainz, Germany }
\date{\today}

\begin{abstract}
Using the momentum dependence of the dressed quark mass and the well-known formulae for the mass dependent quark-loop contribution to the light-by-light scattering insertions, we compute the hadronic light-by-light contribution to the muon anomalous magnetic moment. The relevant momentum running in the quark loop is calculated from the $\pi^0$ exchange contribution to the light-by-light scattering. Special emphasis on the reconstruction of the pseudoscalar transition form factor is made, and the $\pi^0$ contribution to the hadronic light-by-light is, as well, updated. 
\end{abstract}


\maketitle

\section{Introduction}

The anomalous magnetic moment of the muon is one of the most accurately measured quantities in particle physics. Any deviation from its prediction in the Standard Model of particle physics is a very promising signal of new physics.

The present world average experimental value of its deviation from the Dirac value, i.e., $\amu=(g_{\mu}-2)/2$, is given by $\amu^{EXP}=116 592 09.1(6.3)\times10^{-10}$~\cite{Bennett:2002jb,Bennett:2004pv,Bennett:2006fi,PDG2014}\footnote{The quoted number is the recent update in Ref.~\cite{PDG2014} of the original published result from Ref.~\cite{Bennett:2006fi}.}. This impressive result is still limited by statistical errors, and a proposal to measure the muon $(g-2)_{\mu}$ to a precision of $1.6 \times 10^{-10}$ has been submitted to FNAL \cite{Carey:2009zzb}.

At the level of the experimental accuracy, the QED contributions to $\amu$ from photons and leptons alone are very well known. Recently the calculation has been completed up to the fifth order ${\cal O}(\alpha_{em}^5)$ in the fine-structure constant $\alpha_{em}$, giving as result for the QED contribution $11658471.8951(80)\times10^{-10}$ \cite{Aoyama:2012wk}.

The main uncertainties at present in the Standard Model calculation for $\amu$ originate from the hadronic vacuum polarization (HVP) as well as hadronic light-by-light scattering (HLBL) corrections. Without being exhaustive, we show representative estimates and their uncertainties for the QED, HVP, HLBL, and the electroweak (EW) corrections in Table~\ref{SMcont}.

The leading-order (LO) HVP is given by $\sigma(e^+e^- \to hadrons)$ data~ \cite{Aoyama:2012wk,Davier:2010nc} (with errors dominated by experimental uncertainties). Alternatively, the spectral functions of the $\tau \to \nu_\tau + hadrons$ can be used thanks to their relation via isospin symmetry to the isovector  $e^+e^- \to hadrons$~\cite{Davier:2010nc}, but the role of $\tau$ data is still under discussion (see, for instance, Refs.~\cite{Davier:2010nc,Benayoun:2012wc,Benayoun:2015gxa}). A next-to-leading-order estimate (NLO) based on the same $\sigma(e^+e^- \to hadrons)$ data is available~\cite{Hagiwara:2011af}, and even recently a next-to-next-to-leading order (NNLO) calculation has been performed~\cite{Kurz:2014wya}.

The HLBL, entering at NLO, has two representative numbers as well~\cite{Jegerlehner:2009ry,Prades:2009tw} which differ on how the particular subtleties of the calculations are performed (for further details, see for example the discussion in Ref.~\cite{Masjuan:2014rea}).

\begin{table}[hpbt]
\caption{Standard Model contributions to $\amu$. The last row {\bf Total} collects the different boldfaced contributions.}
\begin{center}
\begin{tabular}{ccc}
Contribution  &   \hspace{0.2cm}  Result in $10^{-10}$ units &  \hspace{0.2cm}  Ref.\\[3pt]
\hline
{\bf QED} (leptons) & \hspace{0.2cm}  $11658471.8951\pm 0.0080$ & \hspace{0.2cm}  \cite{Aoyama:2012wk}\\ [2pt]
\hline
{\bf HVP}$^{\mathrm{LO}}(e^+e^-)$  & \hspace{0.2cm}  $692.3\pm4.2$ & \hspace{0.2cm}  \cite{Davier:2010nc}\\
HVP$^{\mathrm{LO}}(e^+e^-)$  & \hspace{0.2cm}  $694.9\pm4.3$ & \hspace{0.2cm}  \cite{Hagiwara:2011af}\\
HVP$^{\mathrm{LO}}(\tau)$  & \hspace{0.2cm}  $701.5 \pm 4.7$ & \hspace{0.2cm}  \cite{Davier:2010nc}\\[2pt]
\hline
{\bf HVP}$^{\mathrm{NLO}}$ & \hspace{0.2cm}  $-9.84\pm0.07$& \hspace{0.2cm}  \cite{Hagiwara:2011af}\\
{\bf HVP}$^{\mathrm{NNLO}}$ & \hspace{0.2cm}  $1.24\pm0.01$& \hspace{0.2cm}  \cite{Kurz:2014wya}\\[2pt]
\hline
{\bf HLBL} & \hspace{0.2cm}  $11.6 \pm3.9$ & \hspace{0.2cm}  \cite{Jegerlehner:2009ry}\\
HLBL &  \hspace{0.2cm} $10.5 \pm 2.5$ & \hspace{0.2cm}  \cite{Prades:2009tw}\\[2pt]
\hline
{\bf EW} & \hspace{0.2cm}  $15.36\pm0.10$ & \hspace{0.2cm} \cite{Gnendiger:2013pva}\\[2pt]
\hline
{\bf Total} &  \hspace{0.2cm} $11659182.6\pm5.7$ \\
\hline
\vspace{-1cm}
\end{tabular}
\end{center}
\label{SMcont}
\end{table}

The existing discrepancy between the experimental value for $\amu$ and its Standard Model prediction stands at about $3\sigma$.

In order to interpret the upcoming new experiment at FNAL, with an anticipated precision of $1.6\times 10^{-10}$, there is an urgent need to improve on both the HVP as well as the HLBL contributions. The accuracy of the HVP contribution depends on the statistical error of the experimental data for the $e^+e^-$ annihilation cross-section into hadrons. With future experiments, in particular at BES-III \cite{Asner:2008nq}, one foresees this error to quantitatively decrease. The HLBL cannot be directly related to any measurable cross section however, and requires the knowledge of Quantum Chromodynamics (QCD) contributions at all energy scales. Since this is not known yet, one needs to rely on hadronic models to compute it. Such models introduce some systematic errors which are difficult to quantify. 

The main motivation of this work is to provide a ballpark prediction for the HLBL scattering based on a duality argument between the hadronic degrees of freedom and the well-known quark loop contribution~\cite{Aldins:1969jz,Aldins:1970id,Laporta:1992pa}. 

\begin{figure}[htbp]
\begin{center}
\includegraphics[width=5cm]{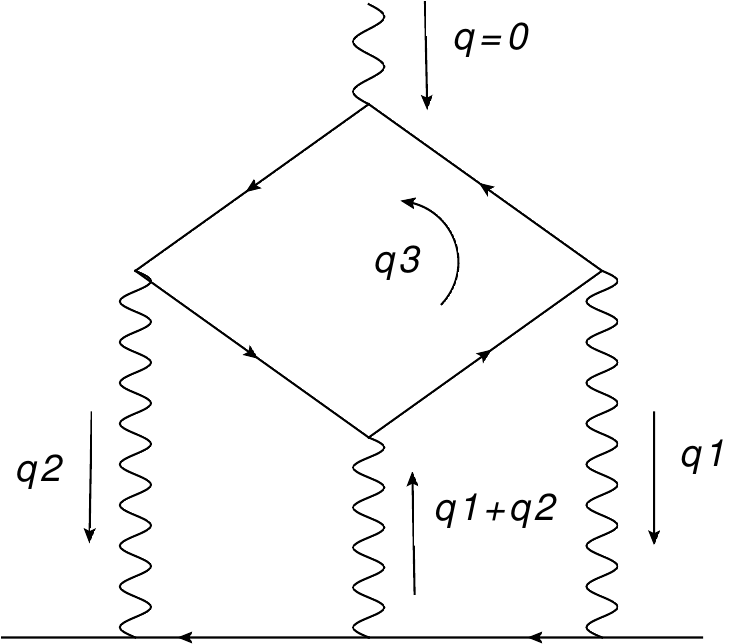}
\caption{Quark loop diagram with running quark mass.}
\label{fig:ql}
\end{center}
\end{figure}

Such a duality estimate can be obtained by invoking a particular regime of QCD where one knows how to perform the calculation responsible for the $\amu^{HLBL}$ (Fig.~\ref{fig:ql}). This is the large-$N_c$ of QCD \cite{tHooft:1973jz,Witten:1979kh} where a quark-hadron duality is accounted for considering that hadronic amplitudes are described by an infinite set of non-interacting and non-decaying resonances. As shown in Ref.~\cite{deRafael:1993za,Prades:2009tw}, the large-$N_c$ limit provides a very useful framework to approach this problem.

\begin{figure}
\begin{center}
\includegraphics[width=8cm]{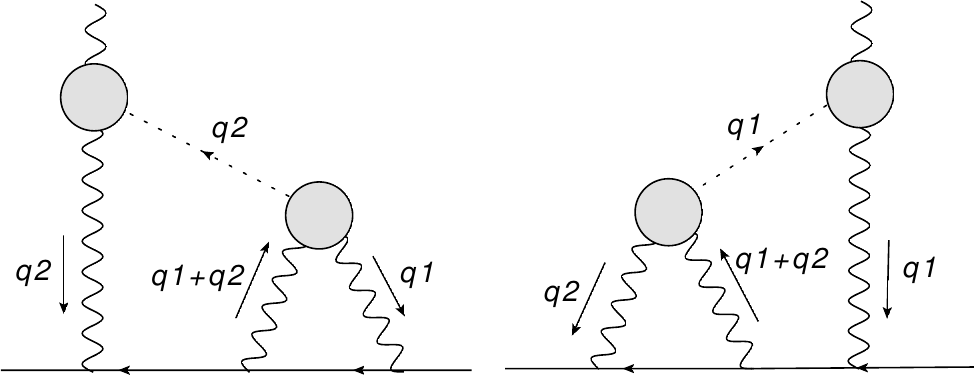}\\[10pt]
\includegraphics[width=4cm]{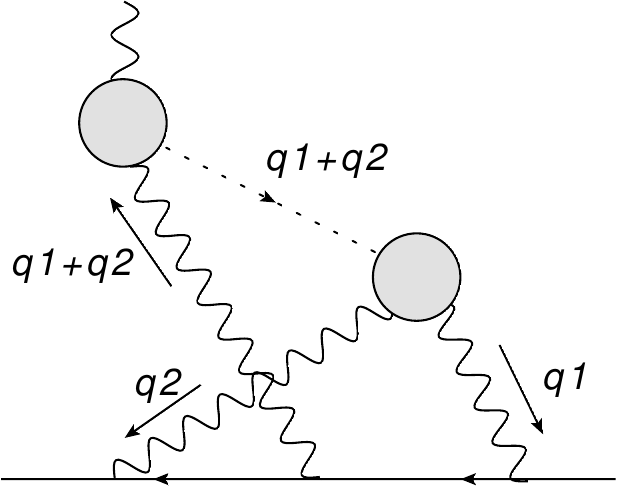}
\caption{Pion-exchange contribution to HLBL scattering. The shaded blobs represent the Form Factor $F_{\pi^0\gamma^*\gamma^*}(q^2_1,q^2_2)$.}
\label{fig:pionexchange}
\end{center}
\end{figure}

Using the large-$N_c$ counting and also the chiral counting, it was proposed in \cite{deRafael:1993za} to split the quark-loop diagram of Fig.~\ref{fig:ql} accounting for the HLBL into a set of different contributions where the numerically dominant contribution arises from the pseudo-scalar exchange diagram shown in Fig.~\ref{fig:pionexchange} \cite{Jegerlehner:2009ry}.

The large-$N_c$ approach however has two shortcomings: firstly, the assumption of pseudoscalar-exchange dominance implies that the remaining pieces are small enough to justify their omission. Although this seems reasonable~\cite{Prades:2009tw}, still the HLBL contribution to $a_{\mu}$ in Table~\ref{SMcont} is larger than the pseudoscalar-exchange one $a_{\mu}^{HLBL;PS}\sim 9\times 10^{-10}$~\cite{Prades:2009tw,Masjuan:2014rea}, and higher excitations may be relevant as well, although neglected so far. This assumption might lead to an underestimation of the final HLBL error. Secondly, calculations carried out in the large-$N_c$ limit demand an infinite set of resonances. As such sum is not known in practice, one ends up truncating the spectral function in a resonance saturation scheme, the so-called Minimal Hadronic Approximation \cite{Peris:1998nj}. The resonance masses used in each calculation are then taken as the physical ones from PDG \cite{PDG2014} instead of the corresponding masses in the large-$N_c$ limit. Both problems might lead to large systematic errors not included so far \cite{Masjuan:2007ay,Masjuan:2012wy}.

It was pointed out in Ref.~\cite{Masjuan:2007ay} that, in the large-$N_c$ framework, the Minimal Hadronic Approximation can be understood from the mathematical theory of Pad\'e Approximants (PA) to meromorphic functions. Obeying the rules from this mathematical framework, one can compute the desired quantities in a model-independent way and even be able to ascribe a systematic error to the approach \cite{Masjuan:2009wy}. One interesting detail from this theory \cite{Queralt:2010sv} is that given a low-energy expansion of a meromorphic function, a PA sequence converges much faster than a rational function with the poles fixed in advance (such as the common hadronic models used so far for evaluating the HLBL), especially when the correct large $Q^2$ behavior is imposed.

The Pad\'e Theory technique here described will be used in the present work for updating the $\pi^0$ exchange contribution to the HLBL accounting, for the first time, for a systematic error. The way such contribution will be derived, in a data-driven approach, provides a simply way to be updated as soon as new experimental data on the $\pi^0$ transition form factor, the $\Gamma(\pi^0 \to \gamma \gamma)$, or experimental information on the doubly virtual form factor will become available. 

This letter is organized as follows. In Section~\ref{S2}, we discuss the way we relate the well-known quark loop~\cite{Aldins:1969jz,Aldins:1970id,Laporta:1992pa} with the $\pi^0$-exchange contribution to the HLBL materialized by the use of a momentum-dependent quark mass function provided in Refs.~\cite{Bhagwat:2003vw,Bhagwat:2007vx}. The duality argument will be used to obtain the averaged momentum running in the quark loop which will allow us to provide a ballpark for the HLBL. The contribution of the $\pi^0$ exchange to the HLBL will be actualized. In Section~\ref{S3} ,we collect our main results together with a discussion of the potential sources of uncertainties in our approach, and  the conclusions.

\section{Quark-hadron duality estimate}\label{S2}

To perform a quark-hadron duality estimate for the HLBL contribution to $\amu$, we now discuss the direct contribution from the quark-loop diagram Fig~\ref{fig:ql}. The diagram with a light-quark running, with a mass of a few MeV, is only valid in the regime where quarks are not confined, so in the perturbative QCD regime. This is not the region dominating the loop integral since we know that the momentum circulating the loop covers the low- and high-energy ranges. From quark models, one can use a constituent quark mass model with mass around $200$ MeV as a rough estimate. Such a value is obtained after comparing the quark model with the experimental value of the HVP~\cite{Pivovarov:2001mw,Erler:2006vu,Boughezal:2011vw,Greynat:2012ww}. This would fail, however, to reproduce the perturbative QCD result at high energies. One, then, desires a running momentum-dependent mass to perform such integration.

\begin{figure}[htbp]
\begin{center}
\includegraphics[width=8.5cm]{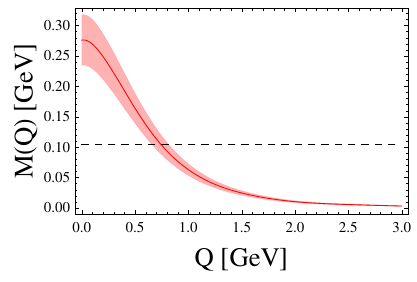}
\caption{Momentum-dependent dressed quark mass,  $M(Q)$, obtained as a solution of the gap equation within the Dyson-Schwinger formalism worked out in Ref.~\cite{Bhagwat:2003vw} extrapolated to the chiral limit~\cite{Bhagwat:2007vx}. The horizontal line represents the muon mass. The band corresponds to a theoretical error of $\pm 15\%$ on top of the chiral limit (solid red line). See main text for details.}
\label{fig:DSE}
\end{center}
\end{figure}
\noindent
Ideally, lattice QCD calculations could be able to calculate, from firsts principles, such momentum-dependent quark masses. As at present lattice QCD calculations are not yet fully feasible for physical pion masses and large volumes, our proposal is to use the dressed-quark mass function $M(Q)$ computed within the Dyson-Schwinger equation (DSE) framework to finally estimate a ballpark prediction for the HLBL (see~\cite{Goecke:2012qm} where a full calculation of the HLBL within DSE is discussed). As shown in Fig.~\ref{fig:DSE}, the solution of the gap equation for a renormalized dressed-quark propagator using DSE provides us with the desired momentum dependent quark-mass function~\cite{Bhagwat:2003vw} which is shown in Fig.~\ref{fig:DSE} after an extrapolation to the chiral limit~\cite{Bhagwat:2007vx}\footnote{We thank the authors of Refs.~\cite{Bhagwat:2003vw,Bhagwat:2007vx} for providing us with their results.}. Such DSE was fitted to unquenched lattice QCD calculations of the quark propagator in the Landau gauge~\cite{Bowman:2002kn,Bowman:2005vx}. Since the results of~\cite{Bhagwat:2003vw,Bhagwat:2007vx} shown in Fig.~\ref{fig:DSE} come from an extrapolation to the chiral limit from the physical light quark mass regime, we add a $\pm 10\%$ as an extrapolation error as suggested in Ref.~\cite{Bowman:2002kn}. On top, since different DSE calculations may yield slightly different quark-mass functions at low energies (see, for instance, a more complete study in Ref.~\cite{Aguilar:2010cn} based on quenched lattice simulations --- but without chiral extrapolation---as well as the discussion in Ref.~\cite{Bowman:2005vx} about the different mass functions from unquenched and quenched lattice studies), and we select Refs.~\cite{Bhagwat:2003vw,Bhagwat:2007vx} as our framework, we add an extra $\pm10\%$ of error to be statistically combined by the one from the chiral extrapolation, resulting in the $\pm15\%$ shown in Fig.~\ref{fig:DSE}.

The full calculation of the quark loop in Fig.~\ref{fig:ql} with full dressed propagators based on DSE is a very difficult and ambitious enterprise (see Ref.~\cite{Goecke:2012qm} for the progress on the field). Such calculation would yield a precise numerical evaluation of that contribution to the HLBL with tiny errors~\cite{Goecke:2012qm}. Our goal here is less ambitious, since we are heading towards a ballpark estimate for the HLBL. To simplify the calculation we notice that if one could assume that all momenta running in the quark loop would be similar, the quark masses would be similar as well~\cite{Pivovarov:2001mw,Erler:2006vu,Boughezal:2011vw,Greynat:2012ww}. If that would be the case, one could then obtain that ballpark estimate using the well-known formulae for the $a_{\mu}^{HLBL}$ for spin$-1/2$ fermions computed in \cite{Aldins:1969jz,Aldins:1970id,Laporta:1992pa}: 

\begin{widetext}
\begin{equation}\label{laporta}
a_{\mu}^{HLBL}(M(Q))=\bigg(\frac{\alpha}{\pi}\bigg)^3 N_c \bigg( \sum_{q=u,d,s} Q_q^4\bigg)\Bigg[ \bigg(\frac32\zeta (3)-\frac{19}{16}\bigg) \frac{m_{\mu}^2}{M(Q)^2} + {\cal O}\bigg(\frac{m_{\mu}^4}{M(Q)^4} \log^2 \frac{m_{\mu}^2}{M(Q)^2} \bigg)\Bigg]\, .
\end{equation}
\end{widetext}

In Eq.~(\ref{laporta}) we display the result up to first order in $(m_{\mu}^2/M(Q)^2)$ but we include all terms up to the fifth order in our numerical calculations~\cite{Jegerlehner:2009ry}. $M(Q)$ represents the three degenerate light quarks, i.e., in a $SU(3)$ symmetric world. Using the same formula, one can account for $SU(3)$ breaking effects by considering a larger strange quark mass. The same applies for including the charm quark in our discussion. These effects will be discussed later on and included in our final results.

To use Eq.~\eqref{laporta}, we need to determine the average momentum $\langle Q \rangle$ that sets the scale of the problem. For that, we will \emph{relate} the hadronic content of the quark loop of Fig.~\ref{fig:ql} onto the dominant hadronic piece of the HLBL. This is the pion-exchange contribution Fig.~\ref{fig:pionexchange}, which is calculable~\cite{Jegerlehner:2009ry,Knecht:2001qg}. This does not mean that the $\pi^0$ saturates the whole HLBL while the rest of the pieces can be neglected, but knowing the $\pi^0$ contribution will be enough for our estimate of the $\langle Q \rangle$.

The $\langle Q \rangle$ to be used in Eq.~\eqref{laporta} is the momentum running in a quark line, and it is in principle unknown. We assume that the photon lines carry also an average momentum $Q_1\sim Q_2\sim Q$. Since the photons in Fig.~\ref{fig:ql} are the same as the ones in Fig.~\ref{fig:pionexchange} with calculable diagrams, we can extract the desired $\langle Q \rangle$ on the photon lines from the $\pi^0$ contribution to HLBL. We assume then, for simplicity (as done in quark-model calculations of the HVP~\cite{Pivovarov:2001mw,Erler:2006vu,Boughezal:2011vw,Greynat:2012ww}), that in average the four quark lines in Fig.~\ref{fig:ql} have the same average momentum $\langle Q \rangle$ as the photon lines with $Q_1,Q_2$. In reality, while the quark lines coupled to the external momentum will indeed have $\langle Q \rangle$, the other two would have $\sim 0$ and $2 \langle Q \rangle$ respectively. For $\langle Q \rangle \sim 0.5$GeV, that would correspond to $M(0)\sim 300$~MeV, and $M(2Q) \sim 50$~MeV which, in average, is $\sim 175$MeV. That corresponds back to $Q\sim0.5$GeV since the mass function behaves linearly at low energies and we profit from it for assuming the four quark lines to have the same momentum. The real and \emph{a priori} assumption in this work is to take the two masses to be equal to their average and substitute them into the quark propagators. To this assumption is difficult to ascribe a theoretical error, though, and we will proceed to consider different average procedures for that purpose. The idea of considering quarks as massive fermions to evaluate the HLBL using Eq.~\eqref{laporta} has been already exploited in Refs.~~\cite{Pivovarov:2001mw,Erler:2006vu,Boughezal:2011vw,Greynat:2012ww}. There, the particular quark mass values were obtained and or crosschecked by the experimental value of the HVP. Here, we proceed differently, namely, asking what averaged photon momenta (leading to the quark mass value) would reproduce the $\pi^0$ exchange contribution to the HLBL. Comparison with previous ballpark estimates~\cite{Pivovarov:2001mw,Erler:2006vu,Boughezal:2011vw}, similar in philosophy, will justify our \emph{a priori} assumption with great success.

The aforementioned calculation yields the average momentum we are interested in by evaluating the first raw momentum about zero of the integrals accounting for the two-loop process in Fig.~\ref{fig:pionexchange}. In this averaging procedure, we consider $Q_1,Q_2$ as well as the combinations $Q=(Q_1+Q_2)/2$ and $Q=\sqrt{Q_1 Q_2}$, and their difference will be ascribed as a theoretical uncertainty.  As we will see later, the statement  $\langle Q \rangle \sim \langle Q_1 \rangle \sim \langle Q_2 \rangle$ is corroborated by the calculation.


The \emph{relation} between the quark loop and the $\pi^0$ exchange we are invoking should be done carefully. The $\pi^0$ contribution to $a_{\mu}^{HLBL}$, the $a_{\mu}^{HLBL;\pi^0}$, has the following structure at low energies~\cite{Knecht:2001qg,Blokland:2001pb}:

\begin{equation}\label{amupi0}
\begin{split}
a_{\mu}^{HLBL;\pi^0}=\bigg(\frac{\alpha}{\pi}\bigg)^3 \frac{N_c^2 m_{\mu}^2}{48 \pi^2 f_{\pi}^2} \left( \log^2\frac{\Lambda_H}{m_{\mu}}+C_1\log\frac{\Lambda_H}{m_{\mu}} +C_0 \right. \\
+{\cal O}\left(m_{\mu}^2/\Lambda_H^2\right) + {\cal O}(\delta) \Big)\, ,
\end{split}
\end{equation}
where $\Lambda_H\gg m_{\mu}$ is a typical hadronic scale, $C_{0,1}$ are constants and $\delta = (m_{\pi}^2-m_{\mu}^2)/m_{\mu}^2$~\cite{Blokland:2001pb}. ${\cal O}(m_{\mu}^2/\Lambda_H^2)$ and ${\cal O}(\delta)$ contain $\log's$ and constant terms~\cite{Blokland:2001pb}.

The $\log^2$ is universal, comes from gauge invariance and the chiral anomaly~\cite{Knecht:2001qf,Knecht:2001qg}, and is not reproduced by the quark-loop calculation~\eqref{laporta}. The $\log$ term is not fixed by symmetries and is related to the divergence of pseudoscalar decay into lepton pairs' loop (inner blob in Fig.~\ref{fig:pionexchange}). In Ref.~\cite{Blokland:2001pb} both the $\log$ and the non-$\log$ terms were analytically calculated after expanding the diagrams in Fig.~\ref{fig:pionexchange} in terms of $m_{\mu}^2/\Lambda_H^2$ and $\delta$ and assuming a particular model for the $\pi^0\gamma^*\gamma^*$ transition form factor. $C_{0,1}$ and the subleading terms are model dependent and provided in~\cite{Blokland:2001pb}.

To illustrate the role of $\Lambda_H$, it is noticed that on the one hand, the pre-factor in Eq.~\eqref{amupi0} can be written in terms of $\Lambda_H$ provided that one can establish a relation with $f_{\pi}$. That relation comes from relizing that the scale is entering in Eq.~\eqref{amupi0} due to the presence of the $\pi^0$-transition form factors in Fig.~\ref{fig:pionexchange}, and yields $\Lambda_H^2=24\pi^2 f_{\pi}^2/N_c$~\cite{Masjuan:2012sk} 
\begin{equation}\label{amupi02}
\begin{split}
a_{\mu}^{HLBL;\pi^0}=\bigg(\frac{\alpha}{\pi}\bigg)^3 \frac{N_c}{2} \frac{m_{\mu}^2}{\Lambda_H^2} \left( \log^2\frac{\Lambda_H}{m_{\mu}}+ \cdots \right) \, . 
\end{split}
\end{equation}

\begin{figure}[htbp]
\begin{center}
\includegraphics[width=8.5cm]{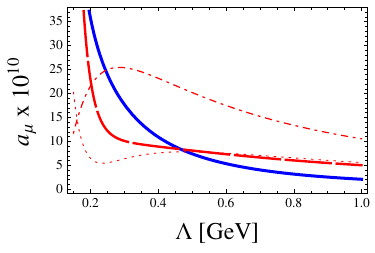}
\caption{Comparison between $\amu^{HLBL}$ calculated using the expansion in Eq.~(\ref{laporta}) up to order $(m_{\mu}^2/M(Q)^2)^{10}$ (solid blue) and the $\pi^0$ contribution to $\amu^{HLBL}$ Eq.~\eqref{amupi02}, in terms of a generic hadronic scale $\Lambda$. The leading $\log^2$ term in~\eqref{amupi02} (dot-dashed red) is compared to the $\log^2+ C_1 \log+ C_0$ pieces (dotted red), and with the result including also ${\cal O}(m_{\mu}^2/\Lambda^2)$ and ${\cal O}(\delta)$ corrections (long-dashed red).}
\label{fig:amuplot2}
\end{center}
\end{figure}

Fig.~\ref{fig:amuplot2} explores how Eq.~\eqref{amupi02} evolves if we artificially vary $\Lambda_H$, represented in that figure by a generic scale $\Lambda$. The $\log^2$ term (dot-dashed line) is pretty much canceled by the rest of the terms in Eq.~(\ref{amupi0}), mainly due to the large coefficient $C_1$ of the $\log$ piece~\cite{Blokland:2001pb,Boughezal:2011vw,Jegerlehner:2009ry}. Such cancelation is illustrated by the dotted line in Fig.~\ref{fig:amuplot2} where the $\log^2$ together with the $C_1 \log$ and $C_0$ pieces are included. The long-dashed line includes also ${\cal O}(m_{\mu}^2/\Lambda_H^2)$ and ${\cal O}(\delta)$ terms. 
The cancellation of the $\log^2$ term for $M>m_{\pi}>m_{\mu}$ was used as an argument in Ref.~\cite{Boughezal:2011vw} to justify that the HLBL can be estimated by the corresponding quark loop function~\eqref{laporta}.

On the other hand, we can introduce a hadronic scale in Eq.\eqref{laporta} as well by defining now $\widetilde{\Lambda_H}=2M(Q)$ and compare Eq.~\eqref{laporta} (blue line) with Eq.~\eqref{amupi02} (red lines) in Fig.~\ref{fig:amuplot2} in terms of a generic scale $\Lambda$ which is introduced for illustrative purposes only since $\Lambda_H \neq \widetilde{\Lambda_H}$. 
The $\Lambda$ variation in Fig.~\ref{fig:amuplot2} should not be attributed to $Q$ since that would imply a $Q$ dependence on $f_{\pi}$ which should not be the case (see, for instance, Ref.~\cite{Pagels:1979hd}), but rather an academic exercise. Indeed, although the behaviors shown are similar, the main difference is the relevant scale for each case ($\Lambda_H \leq 1$GeV for Eq.~\eqref{amupi02}, while $\widetilde{\Lambda_H}=2M(Q) \sim 400$MeV for Eq.~\eqref{laporta}). The fact that both results coincide around $\Lambda \sim 0.5$GeV is just an indication of the different weights of the contributions entering into the HLBL. Since the $\pi^0$ does not completely saturates the full HLBL, we would expect to find the scale in Eq.~\eqref{laporta} smaller than in Eq.~\eqref{amupi02} since $\Lambda_H \neq \widetilde{\Lambda_H}$. (See Ref.~\cite{Boughezal:2011vw} for further discussions along these lines.) 

Notice that while Eq.~\eqref{laporta} contains three flavors which are taken to be degenerate in Fig.~\ref{fig:amuplot2} for simplicity (SU(3) breaking effects will be discussed later), the $\pi^0$ contribution \eqref{amupi0} accounts exclusively for the up and down quarks and knows nothing about the strange. Adding on top the $\eta$ and $\eta'$ contributions one would effectively take the strange quark as well into account~\cite{Escribano:2013kba,Sanchez-Puertas:2014zla}. A naive estimate of SU(3) breaking effects would single out the strange quark in \eqref{laporta} with a higher mass (for instance $M_s \sim 300$ MeV~\cite{Pivovarov:2001mw,Erler:2006vu}), and would imply a decrease in the HLBL prediction of about $\sim 0.8\cdot 10^{-10}$, well inside the thickness of the solid blue line in Fig.~\ref{fig:amuplot2}. Finally, \eqref{laporta} can also be used to account for the charm mass contribution to the HLBL. Taking $M_c=1.2(2)$GeV, $a_{\mu}^{HLBL;c}=0.4(1)\cdot 10^{-10}$.

We will use, instead of a hadronic model for the transition Form Factors (FF), a sequence of Pad\'e approximants~\cite{Masjuan:2007ay} in two variables~\cite{Baker} built up from the low-energy expansion of the $\pi^0$-FF obtained in \cite{Masjuan:2012wy} after a fit to the experimental data \cite{Behrend:1990sr,Gronberg:1997fj,Aubert:2009mc,Uehara:2012ag}, to minimize a model dependence. This is a well-defined data-driven approach. With more experimental data on the $\pi^0$-FF as well as for the $\Gamma_{\pi^0\rightarrow \gamma\gamma}$ the result is straightforwardly updated. Notice, as well, that Pad\'e approximants are well defined from the mathematical theory, and no ambiguity in how to built them exists~\cite{Baker}. The FF is considered to be off-shell (see Refs.~\cite{Melnikov:2003xd,Jegerlehner:2007xe,Dorokhov:2008pw,Jegerlehner:2009ry,Nyffeler:2009tw,Cappiello:2010uy} where this point is addressed). To match the large momentum behavior with short-distance constraints from QCD, calculable using the OPE, we consider the relations obtained in Ref.~\cite{Knecht:2001xc}.

In practice this amounts to use for the FFs (blobs in Fig.~\ref{fig:pionexchange}) the expression: 
\begin{equation}\label{P01}
\begin{split}
&F_{\pi^*\gamma^*\gamma^*}^{P01}(p_{\pi}^2,q_1^2,q_2^2)\, =\, a \frac{b}{q_1^2-b}\frac{b}{q_2^2-b}\big(1+c\, p_{\pi}^2\big)\, ,
\end{split}
\end{equation}
\noindent
where $p_{\pi}=q_1+q_2$ and the free parameters are matched at low energies with the results in \cite{Masjuan:2012wy}: $a$ is fixed by $\Gamma_{\pi^0\rightarrow \gamma\gamma}=7.63(16)$eV from PDG~\cite{PDG2014} which already incorporates the PrimEx Collaboration result \cite{Larin:2010kq}; and $b$ by a matching to the slope $a_{\pi}=0.0324(22)$ \cite{Masjuan:2012wy}. The parameter $c$ characterizes the off-shellnes of the pion and is obtained by imposing, along the lines of the Pad\'e method, that 
\begin{equation}\label{offshell}
\lim_{q\rightarrow \infty} F_{\pi^*\gamma^*\gamma^*}^{P01}(q^2,q^2,0)\, = f_{\pi} \chi /3\, ,
\end{equation}
where $\chi =( -3.3 \pm 1.1)$ GeV$^{-2}$, with an error of $30\%$ as proposed in Refs.~\cite{Jegerlehner:2009ry,Nyffeler:2009tw,Knecht:2001xc}. The results for the averaged $\langle (Q_1+Q_2)/2\rangle$ and $\langle \sqrt{Q_1 Q_2} \rangle$ momenta running in the quark loop in Fig.~\ref{fig:ql} using the FFs of Eq.~(\ref{P01}) are shown in Table \ref{tab:p01}, third column, where in parenthesis we quote the symmeterized errors from the input uncertainties. 
It is difficult to account for an error estimate on our simplified assumption of having equal momenta running in the quark loop (see, however, the justifications in Refs.~\cite{Pivovarov:2001mw,Erler:2006vu,Boughezal:2011vw}). To that respect, we remark that $\langle Q_1 \rangle =0.53(6)$GeV and $\langle Q_2 \rangle =0.48(2)$GeV yielding similar results well within the range shown in Table~\ref{tab:p01}.

We do not average the results in Table~\ref{tab:p01} since that would reduce the errors. We rather prefer to keep both and present them as a window of results. The difference among them is to be understood as a rough estimate of the theoretical uncertainty due to assuming equal momenta running in the quark loop.

The convergence of the PA sequence to a meromorphic function is guaranteed by Pomerenke's theorem \cite{Pommerenke}. The problem is to know how fast this convergence is and also how to ascribe a systematic error on each element of that sequence. For the particular case of a meromorphic function (such as a Green's function in large-$N_c$ QCD), the simplest way of evaluating a systematic error is by comparing the difference between two consecutive elements on the PA sequence \cite{Masjuan:2009wy}.

In our approach to the FF, we evaluate the systematic error by computing a second element on the PA sequence \cite{Masjuan:2009wy} and compare it with the result using Eq.~(\ref{P01}). The second element is:

\begin{equation}\label{P12}
\begin{split}
F_{\pi^*\gamma^*\gamma^*}^{P12}(p_{\pi}^2,q_1^2,q_2^2)\, & = \\
\frac{a+b\,q_1^2}{(q_1^2-d)(q_1^2-e)}&\frac{a+b \,q_2^2}{(q_2^2-d)(q_2^2-e)}\big(1+c\,  p_{\pi}^2\big)\, ,
\end{split}
\end{equation}
\noindent
with five coefficients to be matched with $\Gamma_{\pi^0\rightarrow \gamma\gamma}$, the slope $a_{\pi}$, the curvature of the pion FF $b_{\pi}=1.06(26)\times 10^{-3}$ \cite{Masjuan:2009wy}, with $\chi$ in Eq.~(\ref{offshell}) and the first vector meson resonance $M_{\rho}=0.776(77)$ GeV (where the error in parenthesis is obtained with the {\it half-width rule} method which accounts effectively for $1/N_c$ corrections on meson masses when using PDG values in large-$N_c$ calculations~\cite{Masjuan:2012gc,Masjuan:2012sk,Masjuan:2013xta}). Strictly speaking, one would need the third derivative of the form factor instead of the resonance mass to construct Eq.~(\ref{P12}). Its absence can be substitute by locating the pole to a given value but only for high enough elements on the PA sequence, not for the first one~\cite{Masjuan:2007ay}.

The results for the average momenta running in the quark loop in Fig.~\ref{fig:ql} using the approximant of Eq.~(\ref{P12}) are shown in Table \ref{tab:p01}, second raw, where again the errors are due to the input ones. We assert that the quark masses produced using Fig.~(\ref{fig:DSE}) from our average momenta are in excellent agreement with the constituent quark masses obtained in previous ballpark determinations of the HLBL~\cite{Pivovarov:2001mw,Erler:2006vu,Boughezal:2011vw} based on the experimental value of the HVP. 

Indeed, using the HVP model of~\cite{Pivovarov:2001mw} up to NLO with $M(Q)=0.167(0.181)$ GeV and including the charm quark contribution ($+12\cdot 10^{-10}$ with $M_c=1.2(2)$ GeV), we obtain $a_{\mu}^{HVP}=775(667)\times 10^{-10}$, in a $SU(3)$ flavor symmetric world. The experimental HVP value $a_{\mu}^{HVP}=692(4)\times 10^{-10}$~\cite{Davier:2010nc} seems to prefer $M(Q)=0.181$ GeV. However, if we allow for a $SU(3)$ flavor violation~\cite{Pivovarov:2001mw,Erler:2006vu} with strange quark mass $M_s=0.300$ GeV (basically $M_s=M+120$MeV~\cite{PDG2014}), then $a_{\mu}^{HVP}=692(601)\times 10^{-10}$ for $M(Q)=0.167(0.181)$ GeV, being in excellent agreement with~\cite{Davier:2010nc}. 
Our HVP can be casted as $a_{\mu}^{HVP}=690(90)\times 10^{-10}$, with an error of $13\%$ which includes, effectively, SU(3) breaking effects as well as the uncertainties on the quark mass evaluation. Including an extra error on the $M_s$ of about $\pm 20\%$ would induce a negligible error on the quoted HVP result.
Our method, then, yields as an aside result, the value of the HVP including $SU(3)$ flavor breaking effects.

\begin{table}[htbp]
\caption{Collected results for the average momentum running in the quark loop in Fig.~\ref{fig:ql}, its corresponding mass $M(Q)$ in Fig.~\ref{fig:DSE}, and the $\amu^{HLBL}$ result in accordance to Eq.~(\ref{laporta}), Fig.~\ref{fig:amuplot}, for both $P^0_1$ and $P^1_2$ parameterizations for the $F_{\pi^* \gamma^*\gamma^*}$. Charm-quark contribution $+0.4(2) \times 10^{-10}$ is included in the last column. Errors shown only from input data.}
\begin{center}
\begin{tabular}{ccccc} 

 		&$Q_i$ & $\langle Q_i \rangle$GeV   & $M(Q_i)$GeV \hspace{0.1cm}& $a^{HLBL}_{ \mu} \cdot 10^{10}$  \\[3pt]
\hline			

$P^0_1$\hspace{0.15cm}&$(Q_1+Q_2)/2$\hspace{0.2cm}&	$0.49(6)$ & $0.167(17)$ & $15.7(2.5)$ \\
	&$\sqrt{Q_1\,Q_2}$\hspace{0.2cm}&	$0.44(5)$ & $0.181(14)$ & $13.6(1.8)$ \\
\hline
\hline	
$P^1_2$\hspace{0.15cm}	&$(Q_1+Q_2)/2$\hspace{0.2cm}&       $0.48(9)$ & $0.169(25)$ & $15.9(3.8)$\\ 
	&$\sqrt{Q_1\,Q_2}$\hspace{0.2cm}& $0.44(8)$ & $0.181(23)$ & $14.0(2.8)$\\
\hline
\end{tabular}
\end{center}
\label{tab:p01}
\end{table}%

The similarity of the results obtained within both approximants~(\ref{P01},\ref{P12}) indicates that the low-energy region (up to $1-2$ GeV) dominates the contribution to $\amu^{HLBL}$~\cite{Bijnens:2001cq,Bijnens:2012an}. To evaluate the error on our approximation we look for the maximum of the difference in the region up to $1$ GeV between the $P^0_1$ and $P^1_2$ parameterizations for $F_{\pi^* \gamma^*\gamma^*}$ as explained in Ref.~\cite{Masjuan:2009wy}. Of course, this difference depends on the energy, and grows as the energy increases. At $1$ GeV, the relative difference is about $5\%$~\cite{Escribano:2013kba}, and we take this error as a conservative estimate of the error on the whole low-energy region. We should add this error to the $\amu^{HLBL}$ results shown in Table \ref{tab:p01}.

For comparison with other approaches~\cite{Hayakawa:1997rq,Knecht:2001qf,Bijnens:2001cq,Melnikov:2003xd,Prades:2009tw,Greynat:2012ww,Goecke:2012qm}, we quote what would be the pion-exchange piece to the light-by-light scattering contribution if one would use the parameterizations in Eqs.~(\ref{P01}) and (\ref{P12})  with such purpose. Eq.~(\ref{P01}) yields $\amu^{HLBL;\pi^0} = 6.49(56)\times 10^{-10}$ and Eq.~(\ref{P12}) $\amu^{HLBL;\pi^0} = 6.51(71)\times 10^{-10}$, in good agreement with the result $\amu^{HLBL;\pi^0} = (7.2\pm 1.2)\times 10^{-10}$ obtained in Ref.~\cite{Jegerlehner:2009ry}, but with reduced errors, yielding one of the most important results in this work. For the pion-pole contribution (i.e., when the offshellnes of the pion is swiched-off, $c=0$ in Eqs.~(\ref{P01}) and (\ref{P12})), we would obtain $\amu^{HLBL;\pi^0} = 5.52(27)\times 10^{-10}$ and $5.55(34)\times 10^{-10}$, respectively, in accordance to the result $\amu^{HLBL;\pi^0} = (5.8\pm1.0)\times 10^{-10}$ reported in Ref.~\cite{Knecht:2001qf}.  These results reassure the stability of the PA sequence.

\begin{figure}[htbp]
\begin{center}
\includegraphics[width=8.5cm]{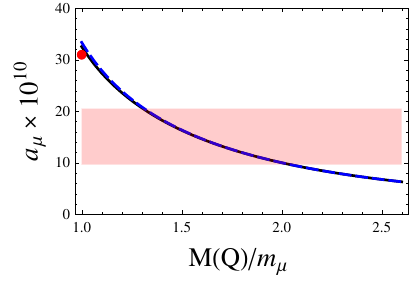}
\caption{$\amu^{HLBL}$ results using the expansion in Eq.~(\ref{laporta}) at order $(m_{\mu}^2/M(Q)^2)^8$ (dashed blue) and at order $(m_{\mu}^2/M(Q)^2)^{10}$ (solid black) in terms of the running quark mass $M(Q)$. The red point indicates the exact result for the point $M(Q)=m_{\mu}$. The horizontal band shows the extremes of the ballpark result from Eq.~(\ref{result}).}
\label{fig:amuplot}
\end{center}
\end{figure}

\section{Reults and Conclusions}~\label{S3}

To quote a final number for $\amu^{HLBL}$ implies to consider several sources of errors: 
\begin{itemize}
\item Firstly, we have the error coming from the experimental inputs used to build our approximants, which arises mainly from the fit to the experimental data on the FF, and the offshellness of the pion. With the new forthcoming experimental data on that FF at BES-III \cite{Asner:2008nq} we intuit lower input errors on our results. 
\item Secondly, we have a $\pm 10\%$ error due to the departure from the chiral limit shown in Fig.~\ref{fig:DSE}. We have a second $\pm 10\%$ of error coming from comparing different DSE studies of the momentum dependence on the running quark mass (see discussion in Sec. II) which yields a combined $\pm 15\%$ error on the final quark-mass determination with a fixed momentum. 
This combined $15\%$ error induces an impact of about $25\%$ on $\amu^{HLBL}$ ($\pm 4.0$ and $\pm 3.4$ in units of $10^{-10}$ for $(Q_1+Q_2)/2$ and $\sqrt{Q_1 Q_2}$ respectively). We are using the DSE to obtain the momentum dependent dressed quark masses. Once the lattice calculations will reach the physical quark mass values, we hope this $15\%$ error will decrease, both from chiral extrapolation and the DSE model dependence.

\item Furthermore, we also have a systematic error from the PA sequence used. We estimate this to be around $5\%$ (see~\cite{Escribano:2013kba} for details on how to obtain such estimation). That implies an error on $\amu^{HLBL}$ about $\pm 0.8$ for $(Q_1+Q_2)/2$ and $\pm0.6$ for $\sqrt{Q_1\, Q_2}$ in units $10^{-10}$. Since both $P^0_1$ and $P^1_2$ parameterizations for $F_{\pi^* \gamma^*\gamma^*}$ give almost the same results for $\amu^{HLBL}$, no extra error due to the difference between them should be included on $\amu^{HLBL}$. We should remark that the FFs employed here, although constrained by the experimental data, do not have the correct behavior when both photon virtualities ($q_1$ and $q_2$) are very large where a behavior $\sim 1/(q_1^2+q_2^2)$ is predicted \cite{delAguila:1981nk,Braaten:1982yp,Brodsky:1989pv,Masjuan:2015lca}. This fact does not affect our calculation since $\amu^{HLBL}$ is very largely dominated by the low-energy region~\cite{Bijnens:2001cq,Bijnens:2012an}.
\item The last source of error considered comes from the evaluation of $\amu^{HLBL}$ using the order $(m_{\mu}^2/M(Q)^2 )^{10}$ in Eq.~(\ref{laporta}) instead of the full result. The difference is so smooth, Fig.~\ref{fig:amuplot}, that no extra error should be considered so far.
\end{itemize} 
Our ballpark estimate lies then in the range:

\begin{equation}\label{result}
\amu^{HLBL}=[13.6(3.9) \div 15.7(4.8)]\times 10^{-10}\, ,
\end{equation}

\noindent
where the error in parenthesis is the combined systematic and input errors as commented above and the two numbers represent the range due to the two momenta considered in our computations in Table~\ref{tab:p01}. They include, as well, the $SU(3)$ breaking effects discussed in Section II. The charm-quark contribution is included using Eq.(\ref{laporta}) setting its mass to be $M_c=1.2(2)$GeV. The numerical window drawn here can be understood as a rough estimate of our theoretical error on the extraction of the quark mass with the method discussed in this work. Combining both results to yield a unique value would result in a smaller error estimation that would underestimate the impact of the assumptions performed. We prefer to keep both numbers and understand its difference as a theoretical uncertainty.

Although different in nature, our ballpark nicely agrees with previous HLBL estimates~\cite{Pivovarov:2001mw,Erler:2006vu,Boughezal:2011vw}. Ref.~\cite{Pivovarov:2001mw} computed the HVP from a quark model and extract the quark mass $M$ from its experimental value, obtaining $M=179(1)$~MeV assuming $SU(3)$ symmetry, and $M=166(1)$~MeV with $SU(3)$ breaking with $M_s=M+180$~MeV. With such a result, ~\cite{Pivovarov:2001mw} used Eq.~\eqref{laporta} up to second order and found $a_{\mu}^{HLBL} = 14.3\cdot 10^{-10}$ (with charm included).

Ref.~\cite{Erler:2006vu} has a similar philosophy than~\cite{Pivovarov:2001mw}, considering quarks like massive leptons and obtaining the HLBL using Eq.~\eqref{laporta}. The way the quark masses were obtained is, however, different. The authors studied how their masses should be to mimic the renormalization group evolution of the QED coupling~\cite{Erler:2004in}, and they obtained $M_u\sim 175$~MeV, $M_d\sim 178$~MeV, $M_s\sim305$~MeV, $M_c\sim1.18$~GeV, and $M_b\sim4.0$~GeV~\cite{Erler:2004in}.  This set of values was crosscheck by calculating several HVP effects and compared with previous results, comparison used to obtain a model error~\cite{Erler:2006vu}. Later on, all this information was used to predict $a_{\mu}^{HLBL} = 13.7^{-0.27}_{+0.15}\times 10^{-10}$, although argued that the meaningful result would be its upper bound, i.e., $a_{\mu}^{HLBL} < 15.9\cdot 10^{-10}$.

Finally, Ref.~\cite{Boughezal:2011vw} considered, as well, that the quark masses entering in the quark loop in Fig.~\ref{fig:ql} can be estimated by requiring that the same quark masses correctly predicts the HVP, assuming again that the quark loop dominates since the chiral $\log^2$ enhancement in Eq.~\eqref{amupi0}, although not reproducible with the quark loop, is numerically not important. Moreover, they argued that such $\log$ enhancement becomes ineffective if the numerical values of the quark masses become comparable to the physical pion mass. They, then, proceed to calculate the HVP with radiative corrections and fitted to experimental data to extract the quark masses in a $SU(3)$ symmetric case (finding $M \sim 200$MeV) and obtained the range $a_{\mu}^{HLBL} =(11.8 - 14.8)\cdot 10^{-10}$, which effectively contains radiative corrections in the $HLBL$.

An immediate conclusion of the discussion above is that the different ballparks yield very similar quark masses, being our results $M \sim 180$~MeV ($SU(3)$ symmetry) and $M\sim 167$~MeV ($SU(3)$ breaking) in perfect agreement. Notice, however, that in our case, as in~\cite{Erler:2006vu}, the HVP is used as a crosscheck of the values of the masses but not for extracting them, and takes the advantage of the justifications provided in~\cite{Pivovarov:2001mw,Erler:2006vu,Boughezal:2011vw} to consider equal quark masses in both HVP and HLBL. Although we could not justify our assumption of all the quark masses having the same average momentum in Fig.~\ref{fig:ql}, the similarity of results makes a compelling cause for our procedure.


With this ballpark estimate and the numbers collected in Table~\ref{SMcont}, we draw a window for the existing discrepancy between the experimental value for the $(g-2)_{\mu}$ and its Standard Model prediction from $3.0\sigma$ to $2.6\sigma$ considering the lower and upper extremes of our ballpark, respectively.

For comparison with approaches that considered a pseudoscalar-pole contribution (i.e, $c=0$ in Eqs.~(\ref{P01}) and (\ref{P12})) instead of a pseudoscalar-exchange, we also report what would be our ballpark estimation for such scenario:

 \begin{equation}\label{resultons}
\amu^{HLBL}=[9.4(2.4) \div 12.9(3.0)]\times 10^{-10}\, .
\end{equation}

In summary, we presented a ballpark estimate of the hadronic light-by-light scattering contribution to the $(g-2)_{\mu}$ based on a duality argument, and estimated the average momentum flowing in the quark loop diagram of Fig.~\ref{fig:ql} from a hadronic parameterization of the $\pi^0$ transition form factor appearing in Fig.~\ref{fig:pionexchange}. This average momentum, then, allowed us to calculate the momentum dependent quark mass in the quark-loop result from $\amu^{HLBL}$. We employed the theory of Pad\'e approximants for evaluating the $\pi^0$ exchange contribution to the HLBL and updated their contribution with, for the first time, a systematic error, $\amu^{HLBL;\pi^0} = 6.49(56)\times 10^{-10}$. Most of the recent phenomenological calculations of the $\amu^{HLBL}$ fall into the range obtained in this work~\cite{Hayakawa:1997rq,Knecht:2001qf,Bijnens:2001cq,Melnikov:2003xd,Prades:2009tw,Kampf:2011ty,Greynat:2012ww,Goecke:2012qm,Roig:2014uja}, including previous ballpark estimates~\cite{Pivovarov:2001mw,Erler:2006vu,Boughezal:2011vw}. The shift of our estimate due to the offshellness of the pseudscalar (compare Eq.~(\ref{result}) with Eq.~(\ref{resultons}) ) suggests further investigations for such kind of effects. Beyond this, we notice that contributions of higher pseudoscalar resonances as well as higher spin resonances not included so far in the HLBL may explain our slightly higher results in Fig.~\ref{fig:amuplot} compared to other determinations. Measurements of two-photon decay widths of $\eta(1295) $, $\pi(1300)$, $\eta(1405)$, $\eta(1475)$, as well as scalar, axial-vector, and tensor states would help along these lines.
\\

{\bf Acknowledgement}

We would like to thank Stan Brodsky, Fred Jegerlehner, and Pablo Roig for helpful discussions and to I. Clo\"et for providing us with the momentum dependent dressed quark mass function shown in Fig.~\ref{fig:DSE}. The present work is supported by the Deutsche Forschungsgemeinschaft DFG through the Collaborative Research Center ``The Low-Energy Frontier of the Standard Model" (SFB 1044).

%

\end{document}